\documentclass[twocolumn,showpacs,preprintnumbers,amsmath,amssymb,prl]{revtex4}

\usepackage{graphicx}
\usepackage{dcolumn}
\usepackage{bm}
\usepackage{psfrag}

\usepackage{latexsym}

\begin{document}

\preprint{}

\title{Quantum spin Hall effect in three dimensional materials:\\
Lattice computation of Z$_2$ topological invariants and its application to Bi and Sb}

\author{Takahiro Fukui}
 \affiliation{Department of Mathematical Sciences, Ibaraki University, Mito
310-8512, Japan}
\author{Yasuhiro Hatsugai}
\affiliation{Department of Applied Physics, University of Tokyo, Hongo,
Tokyo 113-8656, Japan}

\date{\today}

\begin{abstract}
We derive an efficient formula for Z$_2$ topological invariants 
characterizing the quantum spin Hall effect.
It is defined in a lattice Brillouin zone, which enables us to
implement numerical calculations for realistic models even in three
dimensions. 
Based on this, we study the quantum spin Hall effect in 
Bi and Sb in quasi-two and three dimensions
using a  tight-binding model.
\end{abstract}

\maketitle

Quantum spin Hall (QSH) effect 
\cite{KanMel05a,KanMel05b,BerZha05,QWZ05,SSTH05} 
has been attracting much current interest as a
new device of spintronics \cite{MNZ03,Sin04,KMGA04,WKSJ04}. 
It is a topological insulator \cite{Wen89,Hatsugai04,Hatsugai05}
analogous to the quantum Hall (QH) effect,
but it is realized in time-reversal (${\cal T}$) invariant systems.
While QH states are specified by  Chern numbers \cite{TKNN82,Koh85}, 
QSH states are characterized by Z$_2$ topological numbers
\cite{KanMel05b}.

Graphene has been expected to be in the QSH phase \cite{KanMel05a,KanMel05b}. 
However, recent calculations have suggested that 
the spin-orbit coupling in graphene is too small to reveal the QSH
effect experimentally \cite{YYQZF06,MHSSKM06}.
Recently, it has been  pointed out that
Bi thin film is another plausible material for QSH effect 
\cite{Mur06}.
Also by the idea of adiabatic deformation of the diamond lattice, 
it has been conjectured
that Bi in three dimensions (3D) is in a topological phase
\cite{FKM06}.

While systems in two dimensions (2D) are characterized by a single Z$_2$
topological invariant, four independent Z$_2$ 
invariants are needed in 3D
\cite{MooBal06,Roy0607,FKM06}.
This makes it difficult to investigate realistic models,
in which complicated many-band structure is involved.
Therefore, for the direct study of Bi in 3D as well as for the search for
other materials,
to establish a simple and efficient computational method of Z$_2$
invariants in 3D is an urgent issue to be resolved.

In this paper, we present a method of computing Z$_2$ invariants
based on the formula derived by Fu and Kane \cite{FuKan06}
together with the recent development of computing Chern numbers in a lattice
Brillouin zone \cite{FHS05,SWSH06,FukHat06}.
This method is simple enough to compute Z$_2$ invariants  
even for realistic 3D systems. 
Based on this,
we study a tight-binding model for Bi and Sb.

First, we derive a lattice version of the Fu-Kane formula \cite{FuKan06}.
To this end, we restrict our discussions, for simplicity,  
to systems in 2D, where a single Z$_2$ invariant is relevant.  
Let ${\cal T}$ be the time-reversal transformation 
${\cal T}=i\sigma^2K$, and assume that the Hamiltonian in the momentum
space ${\cal H}(k)$ transforms under $\cal T$ as
${\cal T}H(k){\cal T}^{-1}={\cal H}(-k)$.
Let $\psi(k)=(|1 (k)\rangle,\cdots,|2M (k)\rangle )  $ 
denote the  $2M$ dimensional ground state multiplet 
of the Hamiltonian:
${\cal H}(k)|n(k)\rangle=E_n(k)|n(k)\rangle$ 
\cite{Hatsugai04,Hatsugai05}.
Assuming that the many-body energy gap is finite,
we focus on  topological invariants under  the U$(2M)$ transformation
\begin{alignat}{1}
\psi(k) \rightarrow \psi(k) U(k),\ \ U(k)\in \text{U}(2M) .
\label{GauTra}
\end{alignat}
As discussed 
\cite{KanMel05b,FukHat06}, the pfaffian defined by
$p(k)={\rm pf} \Psi ^\dagger ({\cal T} \Psi )$
characterizes the topological phases of ${\cal T}$ invariant systems.
To be precise, the systems belong to topological insulator
if the number of zeros of the pfaffian in half the Brillouin zone 
is 1 (mod 2), and belong to simple insulator otherwise.
This number has been referred to as Z$_2$ invariant.
It should be noted that
under Eq. (\ref{GauTra}), the pfaffian $p(k)$ transforms as
$p(k)\to e^{-i\phi(k)}p(k)$,
where $\phi(k)$ is the U(1) part of U$(2M)$ defined through the relation
$e^{i\phi(k)}\equiv\det U(k)$.

\begin{figure}[htb]
\includegraphics[width=0.99\linewidth]{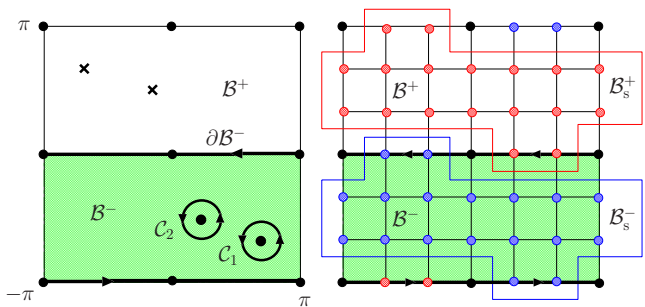}
\caption{Left: The Brillouin zone, where shaded region denotes ${\cal B}^-$. 
The thick lines indicate the boundary of ${\cal B}^-$. 
The integration over ${\cal C}_j$ gives a winding number.
Right: A lattice on the Brillouin zone. The sites in ${\cal B}_{\rm s}^-$,
${\cal B}_{\rm s}^+$, and ${\cal B}_{\rm s}^0$ are denoted by blue, red,
and black circles, respectively. The shaded region denotes the plaquettes 
in ${\cal B}^-$.
}
\label{f:BriZon}
\end{figure}

Recently, Fu and Kane \cite{FuKan06}
have shown that the Z$_2$ invariant
is expressed alternatively by
\begin{alignat}{1}
D&=\frac{1}{2\pi i}\left[
\oint_{\partial{\cal B}^-}A-\int_{{\cal B}^-} F 
\right],
\label{FuKanFor}
\end{alignat} 
where
${\cal B}^-=[-\pi,\pi]\otimes[-\pi,0]$ (See Fig. \ref{f:BriZon}), and where 
$A$ and $F$ is, respectively,  
the Berry gauge potential and associated field strength defined by
$A=
{\rm Tr \,} \psi ^\dagger d \psi$ and $F=dA$
\cite{Hatsugai04,Hatsugai05}.
Notice that the gauge transformation (\ref{GauTra}) yields
$A\to A+i d\phi$.
This implies that the gauge of the Berry gauge potential can be fixed by the
condition that {\it the pfaffian $p(k)$ is real positive.}
Note also that 
$A(-k)=A(k)$ 
holds in this gauge. This is a kind of ${\cal T}$
constraint, as stressed by Fu and Kane \cite{FuKan06}.
The zeros of $p(k)$ thus serve as an obstruction of the gauge fixing
\cite{Roy0604}.

In systems with breaking ${\cal T}$ symmetry like QH effect,
such an obstruction gives in general a nontrivial Chern number.
Contrary to this, in systems under consideration, 
the Chern number always vanishes due to ${\cal T}$ invariance.
Even in this case, an obstruction occurs in the Brillouin zone
as long as the zeros of the pfaffian exist.
Since the zeros occur at the time reversed pairs of points $\pm k_j^*$, 
the vortices around these pairs
are opposite and therefore cancel each other, giving vanishing Chern number.
Nevertheless
the sum of vorticities 
in {\it half the Brillouin zone}, e.g., in $\cal B^-$, 
just gives the number of zeros of $p(k)$ (mod 2).
Imagine, for example, that two zeros exist in ${\cal B}^-$. 
Since they are generically
first order zeros, the winding numbers are $\pm1$. Then
their sum is restricted to $\pm2$ and 0, which can be denoted as ``0 mod 2''.
It thus turns out that 
the Fu-Kane formula (\ref{FuKanFor}) counts the vorticities in half the
Brillouin zone.
So far we have discussed in a specific gauge, but
in any other gauge, $D$ changes by 2,
provided that the gauge keeps ${\cal T}$ constraint.
Therefore, $D$ is indeed a Z$_2$ topological invariant. 
It has also a topological stability against small perturbation as long
as the many-body gap is finite.
As we will show below,
this expression for Z$_2$ invariant is convenient for numerical computations.

Define a lattice on the Brillouin zone,
\begin{alignat}{1}
k_\ell=\pi(j_1/N_1,j_2/N_2), \quad j_\mu=-N_\mu,\cdots,N_\mu.
\end{alignat}
The sites labeled by $k_\ell$
are divided into three sets, ${\cal B}_{\rm s}^\pm$  
and ${\cal T}$ invariant sites ${\cal B}_{\rm s}^0$ 
denoted by red, blue and 
black circles in Fig. \ref{f:BriZon}, respectively.
Here, $\cal T$ invariant sites are specified by the property that 
${\cal T}H(k_\ell){\cal T}^{-1}=H(k_\ell)$.
As a  ${\cal T}$ constraint,
we choose the states at $-k_\ell\in{\cal B}_{\rm s}^+$ 
as their Kramers doublets at $k_\ell\in{\cal B}_{\rm s}^-$.
Suppose that at $k_\ell$ 
the spectrum is arranged as
$\varepsilon_n(k_\ell)\le\varepsilon_{n+1}(k_\ell)$.
Then the states at $-k_\ell$ 
can be constrained as
\begin{alignat}{1}
|n(-k_\ell)\rangle={\cal T}|n(k_\ell)\rangle, \quad \mbox{for} 
\quad k_\ell\in{\cal B}_{\rm s}^- .
\label{LatTimRevConBP}
\end{alignat}
On the other hand, both of the Kramers doublets are included in
${\cal B}_{\rm s}^0$: The spectrum in this set can be arranged in
general as 
$\varepsilon_{2n-1}(k)=\varepsilon_{2n}(k)\le\varepsilon_{2n+1}(k)\cdots$.
Therefore, we enforce the constraint
\begin{alignat}{1}
|2n(k_\ell)\rangle={\cal T}|2n-1(k_\ell)\rangle, 
\quad \mbox{for}\quad k_\ell\in{\cal B}_{\rm s}^0 .
\label{LatTimRevConBTR}
\end{alignat}
With these constrained states, we define a link variable
\begin{alignat}{1}
U_\mu(k_\ell)={\cal N}_\mu^{-1}(k_\ell)\det  
\psi^\dagger(k_\ell)\psi(k_\ell+\hat\mu) ,
\label{LinVar}
\end{alignat}
where ${\cal N}_\mu^{-1}(k_\ell)=|\det  
\psi^\dagger(k_\ell)\psi(k_\ell+\hat\mu)|$,
and associated field strength through a plaquette variable
\begin{alignat}{1}
F_{12}(k_\ell)&=
\ln U_1(k_\ell)U_2(k_\ell+\hat 1)U_1^{-1}(k_\ell+\hat 2)U_2^{-1}(k_\ell) ,
\end{alignat}
where $F_{12}$ is defined within the branch
$F_{12}/i\in(-\pi,\pi)$.

The sum of $F_{12}$ over ${\cal B}^-$ can be written as a similar
formula to Eq. (\ref{FuKanFor}). To see this, it is convenient to define
a gauge potential via
$A_\mu(k_\ell)=\ln U_\mu(k_\ell)$ 
also in the branch $A_\mu(k_\ell)/i\in(-\pi,\pi)$.
Then the field strength can be rewritten as
\begin{alignat}{1}
F_{12}(k_\ell)
&=\Delta_1 A_2(k_\ell)-\Delta_2A_1(k_\ell)+2\pi i n_{12}(k_\ell) ,
\end{alignat}
where {\it integral} field $n_{12}(k_\ell)$ has been introduced so as to 
match the branches of both sides \cite{Luescher99,FSW01,FHS05}.
Thus, we reach 
\begin{alignat}{1}
\sum_{k_\ell\in {\cal B}^-}F_{12}(k_\ell)=
\sum_{k_\ell\in\partial{\cal B}^-}A_1(k_\ell)
+2\pi i\sum_{k_\ell\in{\cal B}^-}n_{12}(k_\ell) ,
\end{alignat}
where the sums of $F_{12}$ and of $n_{12}$ are over the plaquettes in the
shaded region denoted by ${\cal B}^-$ in Fig. \ref{f:BriZon}. 
The sum of $A_\mu$ is over the links of the
boundary of ${\cal B}^-$ specified by thick lines in Fig. \ref{f:BriZon}.
Therefore, a lattice version of $D$ is
\begin{alignat}{1}
D_{\rm L}
&\equiv\frac{1}{2\pi i}\left[
\sum_{k_\ell\in\partial{\cal B}^-}A_1(k_\ell)
-\sum_{k_\ell\in{\cal B}^-}F_{12}(k_\ell)
\right]
\nonumber\\
&=-\sum_{k_\ell\in{\cal B}^-}n_{12}(k_\ell) .
\end{alignat}
This formula for the Z$_2$ invariants is one of the main results
of this paper. Indeed this formula has the following desired properties.
Firstly, it is strictly integral.
Secondly, though the ground state multiplet can be mixed by 
Eq. (\ref{GauTra}), it is SU$(2M)$ invariant. 
Finally, it changes by 2 under the remaining U$(1)$ transformation, and
hence, it is Z$_2$ invariant. The last property will be proved
elsewhere, though it is not difficult.

In 3D, it has been shown that the phases of ${\cal T}$ invariant systems
are classified by four independent Z$_2$ invariants
 \cite{MooBal06,Roy0607,FKM06}.
To compute them, let us define six two-dimensional tori, 
according to Moore and Balents \cite{MooBal06}. 
For example, fix the third momentum to $k_3=0$ or $\pi$, then we have 
two tori spanned by $k_1$ and $k_2$ which we denote $Z_0$ and $Z_\pi$
torus, respectively. 
Applying the previous techniques, 
we can compute two Z$_2$ invariants $D_{\rm L}$ which are referred to as 
$z_0$ and $z_\pi$. In the same way, we have six invariants 
$x_0$, $x_\pi$, $y_0$, $y_\pi$, $z_0$, and $z_\pi$ living on six tori
$X_0$,  $X_\pi$,  $Y_0$,  $Y_\pi$, $Z_0$,  and $Z_\pi$, respectively.    
There are two constraints, however:
$x_0x_\pi=y_0y_\pi=z_0z_\pi$ (mod 2),
and therefore, four invariants among six are independent \cite{MooBal06}.
According to Fu {\it et al.} \cite{FKM06}, we choose them as 
$\nu_0=x_0x_\pi$, $\nu_1=x_\pi$, $\nu_2=y_\pi$, and $\nu_3=z_\pi$, and
denote them as $\nu_0;(\nu_1\nu_2\nu_3)$.
As is known in the QH effects, non-trivial structures of topological
ordered states
are hidden in the bulk and play  physical roles near the boundaries 
as chracteristic edge states \cite{Hat93}.
Based on the principle,
by investigating the relationship
between the Z$_2$ invariants and surface states, 
Fu {\it et al.} have clarified
that there are basically three phases;
simple band insulator,
weak topological insulator (WTI) which is topological but
weak against disorder, 
and more robust strong topological insulator (STI) \cite{FKM06}.

\begin{figure}[hbt]
\begin{tabular}{cccc}
\includegraphics[width=0.25\linewidth]{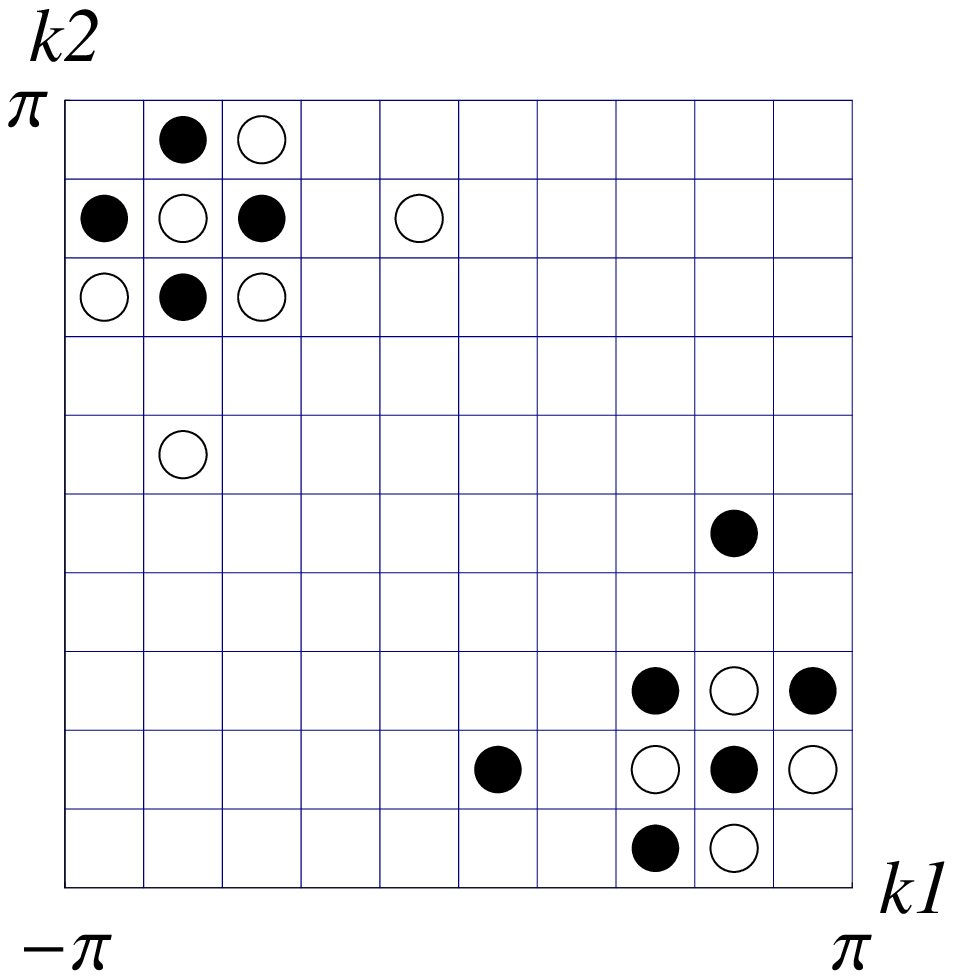}
&
\hspace*{-3mm}
\includegraphics[width=0.25\linewidth]{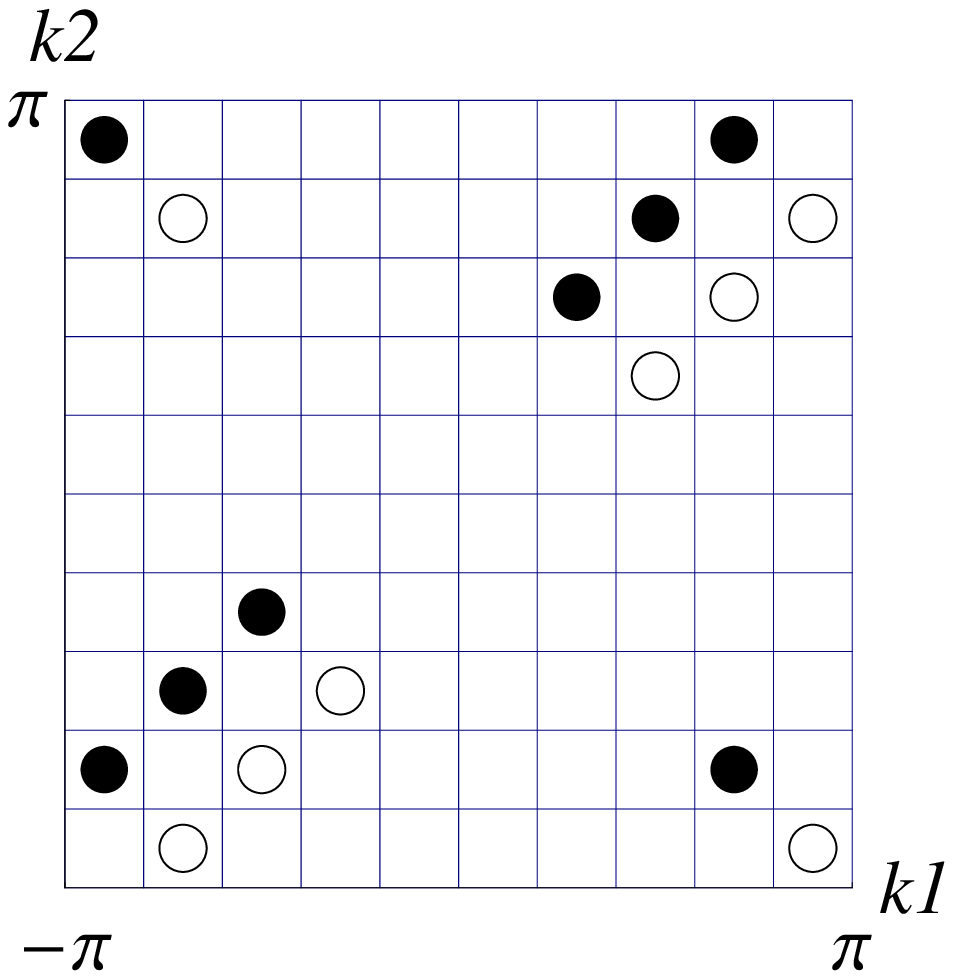}
&
\hspace*{-3mm}
\includegraphics[width=0.25\linewidth]{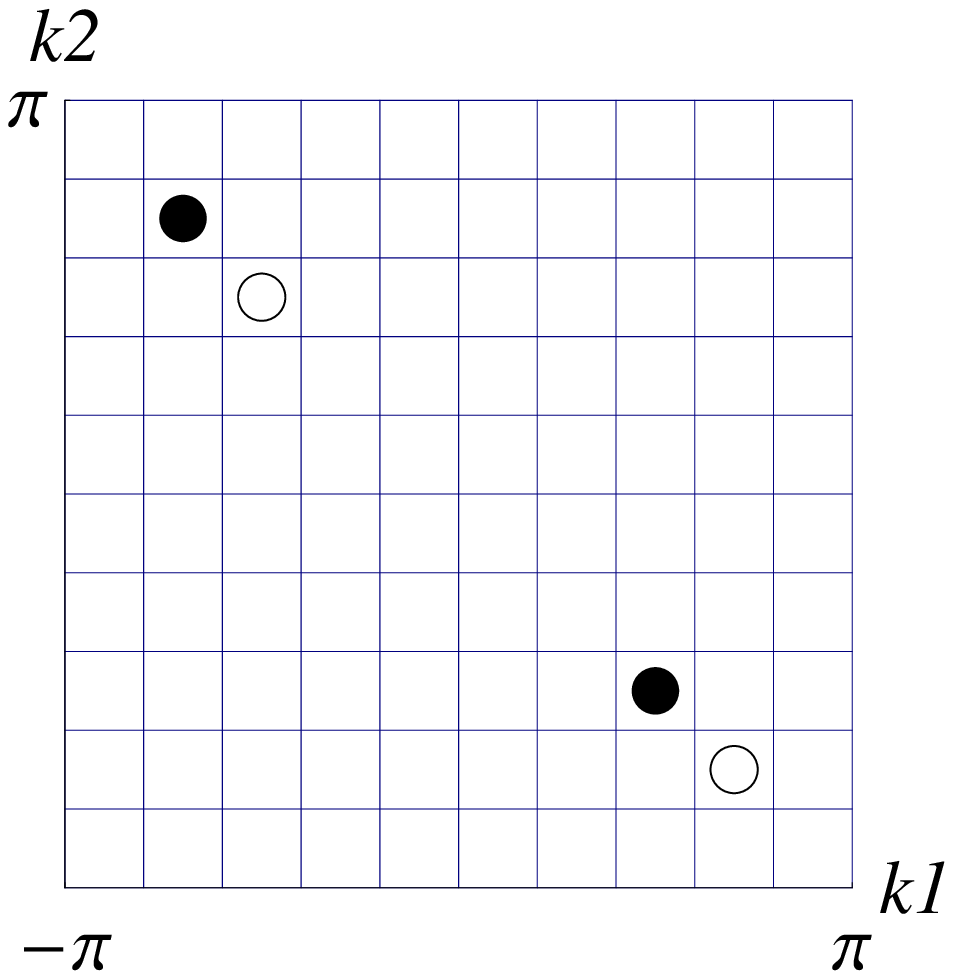}
&
\hspace*{-3mm}
\includegraphics[width=0.25\linewidth]{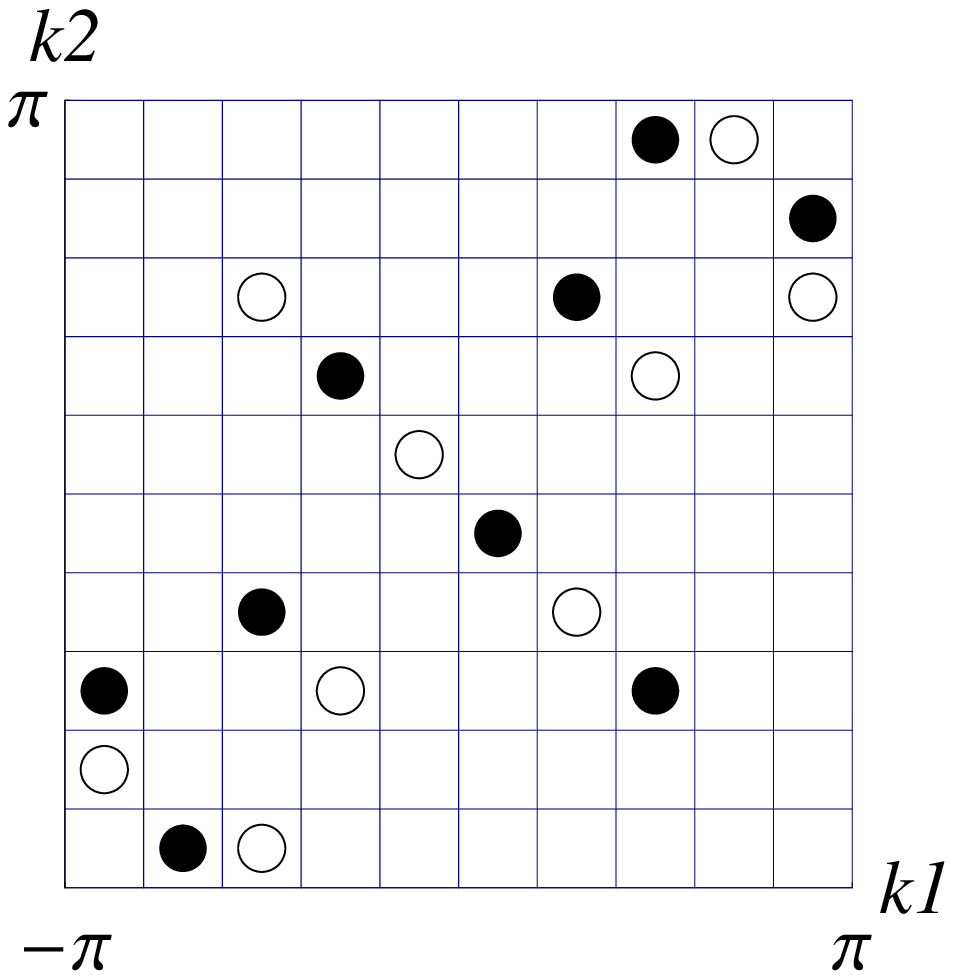}
\end{tabular}
\caption{The $n$-field configurations for Bi (left two panels)
and Sb (right two panels) 
computed by the gauge that the pfaffian is real positive.
The first and third (second and fourth) show the configurations
on the $Z_0$ ($Z_\pi$) torus. 
The white and black circles denote $n_{12}=1$ and $-1$, respectively, 
while the blank denotes 0. 
These read $z_0=2$ and $z_\pi=0$ for Bi, and 
$z_0=0$ and $z_\pi=1$ for Sb. 
}
\label{f:BiSb}
\end{figure}

Recently, Murakami \cite{Mur06} has pointed out the possibility of QSH effect 
in Bi. Though Bi is a semimetal, the valence band and conduction band 
keep the direct gap throughout the Brillouin zone. 
Fu {\it et al.} have studied 
solvable tight-binding  models with the diamond structure, and
predicted that the valence band of Bi is characterized by the
WTI phase specified 0;(111), based on the observation that 
the structure of Bi 
can be viewed as an adiabatically distorted cubic lattice toward 
the diamond lattice.
However, since a realistic tight-binding model including $s$ and $p$
orbitals with nearest neighbor, second neighbor, and third
neighbor hoppings has indeed complicated band structure, 
we  calculate the Z$_2$ invariants directly for 
heavy group V elements.

\begin{figure}[htb]
\begin{tabular}{cc}
\hspace*{-3mm} 
\includegraphics[width=0.47\linewidth]{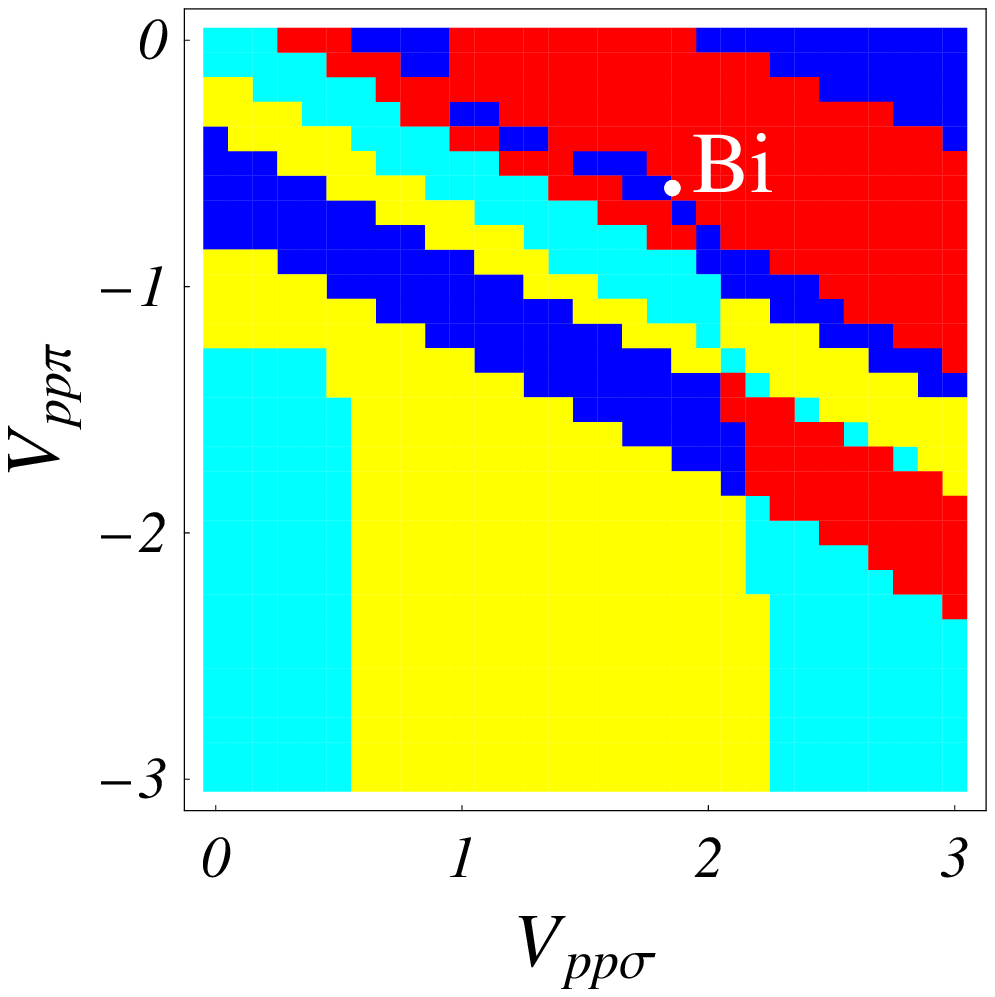}&
\includegraphics[width=0.47\linewidth]{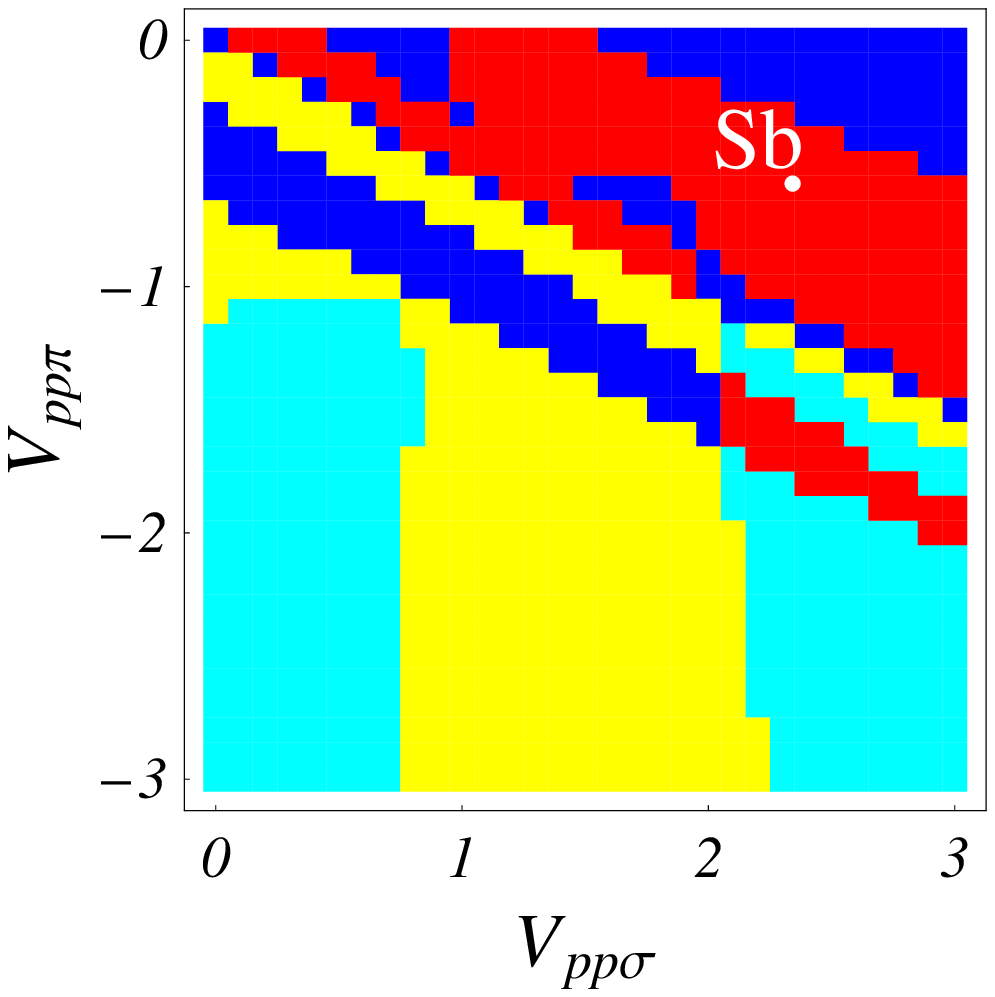}
\end{tabular}
\caption{
Phase diagrams of Bi (left) and Sb (right)
as functions of  $V_{pp\sigma}$ and $V_{pp\pi}$.
Other parameters are the same as those in Ref. \cite{LiuAll95}.
The colored regions denote different phases: 
red for 1;(111), yellow for 1;(000), 
lightblue for 0;(111), and blue for 0;(000).
The white points indicate the position of the parameters for Bi
 and Sb in Ref. \cite{LiuAll95}.}
\label{f:PhaDia}
\end{figure}

In what follows,
we investigate real materials by the tight-binding models in Ref. \cite{LiuAll95}. 
We first discuss the phase of Bi which is attracting much interest.
We show in Fig. \ref{f:BiSb} examples of the
$n$-field configuration computed for Bi. 
Though these calculations are for
rather coarse $10\times10$ lattice, 
we have checked that finer ones indeed
reproduce the same Z$_2$ invariants and our formula is actually
convergent even in this mesh size. For Bi, 
these figures tell that $z_0=z_\pi=0$ mod 2. 
The other Z$_2$ invariants are also 0 mod 2, and it turns out that 
the valence bands of Bi in 3D is specified by 0;(000) phase.
This result is contradictory to the conjecture by 
Fu {\it et al.} mentioned-above.
This suggests that along adiabatic distortion of the lattice,
some topological changes should occur.
Indeed, a slight change of the Slater-Koster parameters can lead to 
different phases. 
Among adjustable 14 parameters \cite{LiuAll95},
crucial ones may be $V_{pp\sigma}$ and $V_{pp\pi}$,
nearest neighbor hopping parameters between $p$ orbitals. 
We show in Fig. \ref{f:PhaDia} the phase diagram of Bi 
to discuss the stability of the phase.
This diagram tells that Bi is located in
a small island of 0;(000) phase embedded in 1;(111) phase.
We also understand this feature from a small direct gap of Bi,
12 meV, at the $L$ point \cite{LiuAll95}. 
With varying the parameters, this gap soon
closes and the phase of Bi changes from trivial phase into STI phase. 
We conclude that Bi in 3D dose not show the QSH effect, though it
locates quite near the phase boundary with STI.

\begin{figure}[htb]
\begin{tabular}{c}
\includegraphics[width=0.8\linewidth]{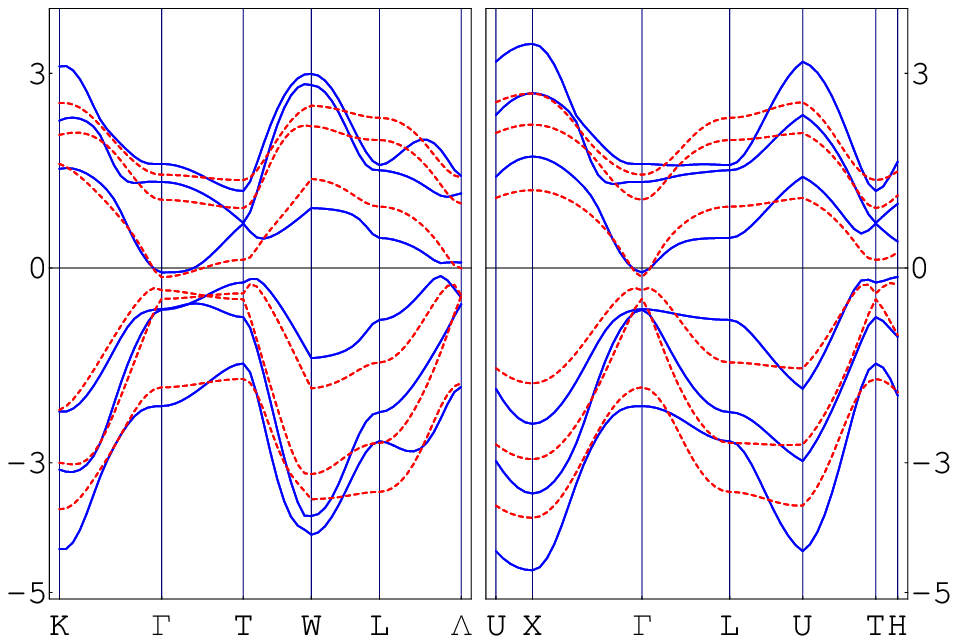}
\vspace*{-5mm}\\
\includegraphics[width=0.8\linewidth]{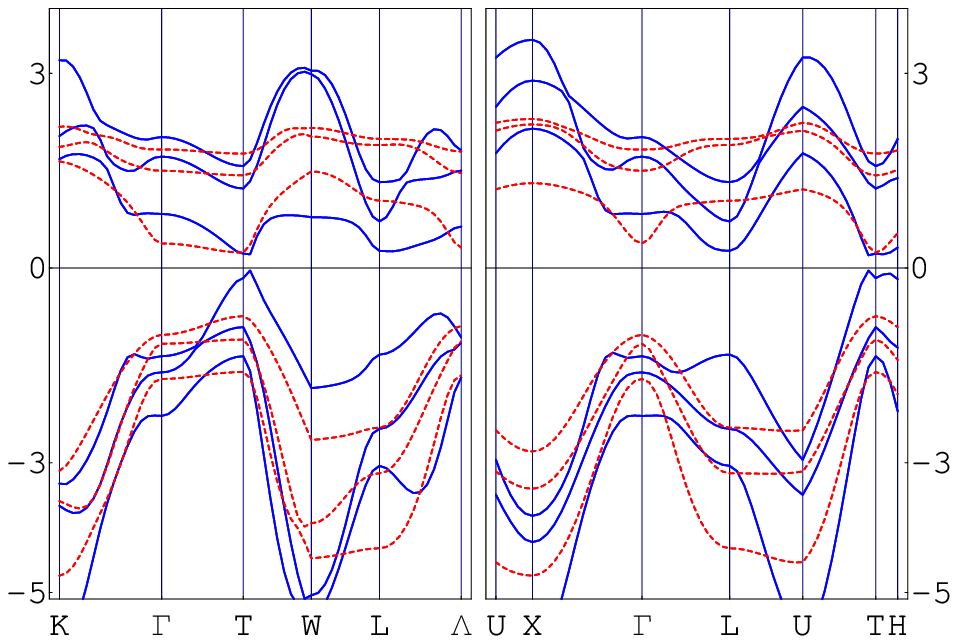}
\end{tabular}

\caption{
Band structure of Bi (upper) and Sb (lower).
Bi: The blue lines are energies at $f=0.5$ 
and the red-dotted lines are at $f=0.1$. 
Sb: The blue lines at $f=0.7$ 
and the red-dotted lines at $f=0.1$. 
There are other bands around $-10$ eV associated mainly with 
$s$ orbitals which are omitted in these figures.
}
\label{f:BanStr2D3D}
\end{figure}

On the other hand, decreasing the thickness, a semimetal-semiconductor
transition occurs, and Murakami has suggested that Bi thin film would be
in QSH phase \cite{Mur06}. 
To study the quasi 2D systems, and also
to clarify the discrepancy between our result and 
the conjecture by Fu {\it et al.}, 
we next investigate the effects of
dimensionality on the present model.
We replace the Slater-Koster parameters $V'_\alpha$
($\alpha=ss\sigma,sp\pi,pp\sigma,pp\pi$) 
for second neighbor hopping in Ref. \cite{LiuAll95}
by $fV'_\alpha$ with a uniform factor $f$. 
This factor $f$ can effectively control the
dimensionality into [111] direction,  interpolating between
Murakami's model at $f=0$ and the 3D model at $f=1$.

We show in Fig. \ref{f:BanStr2D3D} the band structure of Bi with
$f=0.5$ and $0.1$. At $f=1$ (See Ref. \cite{LiuAll95}), 
the overlap energy (indirect gap) 
is $\Delta E=-12$ meV.
With decreasing $f$, 
a semimetal-semiconductor transition occurs at $f\sim0.99$. 
Fig. \ref{f:BanStr2D3D} confirms that 
Bi is indeed a semiconductor at 
$f=0.5$ and $0.1$ whose overlap energy is  
$\Delta E=56$ and $90$ meV, respectively. 
\begin{figure}[htb]
\begin{tabular}{cc}
\includegraphics[width=0.47\linewidth]{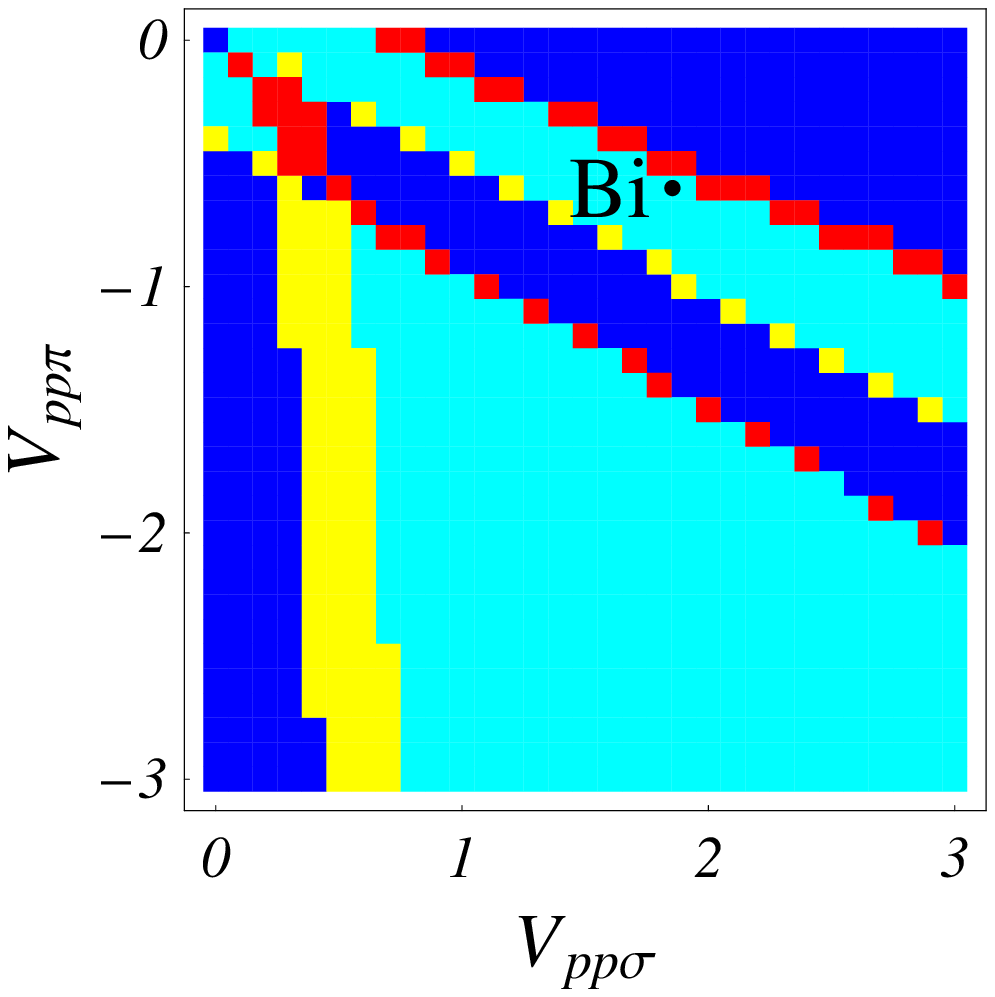}&
\includegraphics[width=0.47\linewidth]{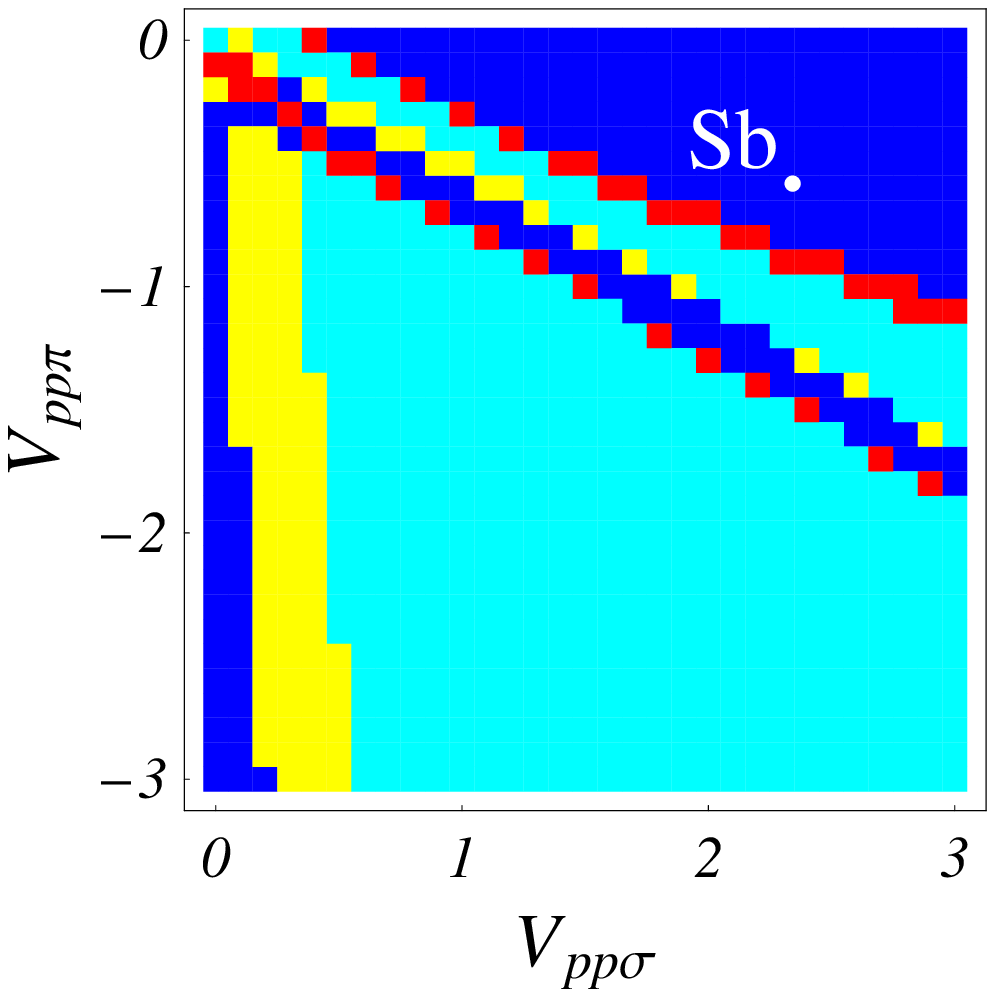}
\end{tabular}
\caption{
Phase diagrams of Bi (upper panel)  and Sb (lower panel)
for $f=0.1$ as functions of  $V_{pp\sigma}$ and $V_{pp\pi}$.
}
\label{f:PhaDia2D3D}
\end{figure}
Near $f=0.993$, the  phase of Bi changes from 0;(000)
into 1;(111). 
With further decreasing $f$ and enhancing the two-dimensionality,  
topological change occurs again near $f\sim 0.223$, and the system
becomes 0;(111), as shown in Fig. \ref{f:PhaDia2D3D}. 
This is just the phase predicted by 
Fu {\it et al. }\cite{FKM06}. 
Therefore, we suggest that the adiabatic distortion of the diamond lattice 
leads to Bi thin film, and along the change of the
dimensionality, the adiabatic distortion does not work, giving rise to
gap-closing and resultant topological changes.
We also conjecture that STI phase is very stable along the change
of $f$, and could be observed by experiments.

Sb is also a semimetal with a larger gap at the $L$ point \cite{LiuAll95}. 
We show in Fig. \ref{f:PhaDia} the phase diagram of Sb. 
It turns out that Sb belongs to the 1;(111) phase even in 3D.
Its location is far from the phase boundary with 0;(000) and therefore, 
it is rather stable. 
We show in Fig. \ref{f:BanStr2D3D} the band structure of Sb. 
With decreasing $f$, a semimetal-semiconductor transition also occurs
at $f\sim0.89$.
Along the change of $f$, topological change occurs once:
Near $f\sim 0.54$, the phase changes from 1;(111) into 0;(000), 
and no 0;(111) phase is observed throughout.
Therefore, the phase of Sb thin film is 0;(000), different from Bi. 
In Fig. \ref{f:PhaDia2D3D}, we show the phase diagram of Sb for
$f=0.1$. 
However, it should be stressed that 
with appropriate thickness, $0.54\leq f\leq0.89$, 
Sb is a semiconductor in STI phase and hence should show QSH effect.

Finally, we comment on
the relationship between the method presented in
this paper and the previous one in Ref. \cite{FukHat06}.
While in the present calculation link variables are
defined with respect to the momentum,
the previous calculation has been implemented 
with respect to twist angles by imposing a
spin-dependent twisted boundary condition.
For systems with appropriate strength of spin-orbit coupling, 
the present computation is more efficient, but for systems with 
very small spin-orbit coupling as well as with inversion symmetry, 
the previous computation by the use of the twisted boundary condition 
gives more reliable results. 
In this sense, both methods are complementary to each other.
Details will be published elsewhere.

We also mention that  recently Fu and Kane \cite{FuKan06b} 
have reached the similar conclusion of the phases of Bi and Sb 
by making the  use of inversion symmetry of the system.
We stress here that our method can apply to any systems, even
without inversion symmetry.

This work was supported in part 
by Grant-in-Aid for Scientific Research  
(Grant No. 17540347, No. 18540365) from JSPS
and on Priority Areas (Grant No.18043007) from MEXT.
YH was also supported in part by the Sumitomo foundation.


\begin{thebibliography}{99} 
\bibitem{KanMel05a}
C.L. Kane and E.J. Mele
Phys. Rev. Lett. {\bf 95}, 226801 (2005). 
%
\bibitem{KanMel05b}
C.L. Kane and E.J. Mele
Phys. Rev. Lett. {\bf 95}, 146802 (2005). 
%
\bibitem{BerZha05}
B. A. Bernevig and S.-C. Zhang,
cond-mat/0504147.
%
\bibitem{QWZ05}
X.-L. Qi, Y.-S. Wu, and S.-C. Zhang,
cond-mat/0505308.
%
\bibitem{SSTH05}
L. Sheng, D.N. Sheng, C.S. Ting, and F.D.M. Haldane,
cond-mat/0506589.
%
\bibitem{MNZ03}
S. Murakami, N. Nagaosa, and S.-C. Zhang,
Science {\bf 301}, 1348 (2003);
%
Phys. Rev. Lett. {\bf 93}, 156804 (2004).
%
\bibitem{Sin04}
J. Sinova, D. Culcer, Q. Niu, N.A. Sinitsyn, T. Jungwirth,
and A.H. MacDonald, 
Phys. Rev. Lett. {\bf 92}, 126603 (2004).
%
\bibitem{KMGA04}
Y.K. Kato, R.C. Myers, A.C. Gossard, and D.D. Awschalom, 
Science, {\bf 306}, 1910 (2004).
%
\bibitem{WKSJ04}
J. Wunderlich, B. K\"astner, J. Sinova, and T. Jungwirth,
Phys. Rev. Lett. {\bf 94}, 047204 (2005).
%
\bibitem{Wen89}
X.G. Wen,
Phys. Rev. B {\bf 40}, 7387 (1989).
\bibitem{Hatsugai04}
Y. Hatsugai,
J. Phys. Soc. Jpn. {\bf 73}, 2604 (2004).
%
\bibitem{Hatsugai05}
Y. Hatsugai,
J. Phys. Soc. Jpn. {\bf 74}, 1374 (2005).
%
\bibitem{TKNN82}
D.J. Thouless, M. Kohmoto, M.P. Nightingale, and M. den Nijs,
Phys. Rev. Lett. {\bf 49}, 405 (1982).
%
\bibitem{Koh85}
M. Kohmoto, 
Ann. Phys. {\bf 160}, 355 (1985).
%
\bibitem{YYQZF06}
Y. Yao, F. Ye, X.-L. Qi, S.-C. Ahang, and Z. Fang,
cond-mat/0606350.
%
\bibitem{MHSSKM06}
H. Min, J.E. Hill, N.A. Sinitsyn, B.R. Sahu, 
L. Kleinman, and A.H. MacDonald,
cond-mat/0606504.
%
\bibitem{Mur06}
S. Murakami,
cond-mat/0607001.
%
\bibitem{FKM06}
L. Fu, C.L. Kane, and E.J. Mele,
cond-mat/0607699.
%
\bibitem{MooBal06}
J.E. Moore and L. Balents,
cond-mat/0607314.
%
\bibitem{Roy0607}
R. Roy,
cond-mat/0607531.
%
\bibitem{FuKan06}
L. Fu and C.L. Kane,
cond-mat/0606336.
%
\bibitem{FHS05}
T. Fukui, Y. Hatsugai, and H. Suzuki,
J. Phys. Soc. Jpn. {\bf 74}, 1674 (2005).
%
\bibitem{SWSH06}
D.N. Sheng, Z.Y. Weng, L. Sheng, and F.D.M. Haldane,
cond-mat/0603054.
%
\bibitem{FukHat06}
T. Fukui and Y. Hatsugai,
cond-mat/0607484.
%
\bibitem{Roy0604}
R. Roy,
cond-mat/0604211.
%
\bibitem{Hat93}
Y. Hatsugai, 
Phys. Rev. Lett. {\bf 71}, 3697 (1993).
%
\bibitem{Luescher99}
M. L\"uscher,
Nucl. Phys. B {\bf 549}, 295 (1999).
%
\bibitem{FSW01}
T. Fujiwara, H. Suzuki, and K. Wu,
Prog. Theor. Phys. {\bf 105}, 789 (2001).
%
\bibitem{LiuAll95}
Y. Liu and R.R. Allen,
Phys. Rev B {\bf 52}, 1566 (1995).
%
\bibitem{FuKan06b}
L. Fu and C.L. Kane,
cond-mat/0611341.
\end{thebibliography}
\end{document}